\documentclass[sigconf]{acmart}
\usepackage[english]{babel}
\usepackage{booktabs}
\usepackage{array}
\usepackage{rotating}
\usepackage{multirow}
\usepackage{booktabs}
\usepackage{xspace}
\usepackage{xcolor}       
\usepackage{colortbl}     

\definecolor{rowko}{RGB}{255,245,245}    
\definecolor{rowpillar}{RGB}{240,246,255} 
\definecolor{rowirp}{RGB}{220,240,255}    
\definecolor{proposed}{RGB}{220,240,255}   
\definecolor{header}{RGB}{240,244,248}      
\newcommand{\cmark}{\textcolor{green!50!black}{\checkmark}}

\renewcommand\footnotetextcopyrightpermission[1]{} 

\setcopyright{none}

\acmConference{40th Brazilian Symposium on Software Engineering}{2026}{São Paulo, SP, Brazil}

\AtBeginDocument{
    
}

\begin{document}

\title{Operationalizing Regulations into Code: A Model to Enhance Governance and Compliance in LLM Selection for Software Engineering}

\author{Jonysberg Quintino}
\affiliation{
  \institution{Centro de Informática - CIn, UFPE}
  \city{Recife}
  \country{Brasil}
}
\email{jpq@cin.ufpe.br}

\author{Hermano Moura}
\affiliation{
  \institution{Centro de Informática - CIn, UFPE}
  \city{Recife}
  \country{Brasil}
}
\email{hermano@cin.ufpe.br}

\author{Filipe Calegário}
\affiliation{
  \institution{Centro de Informática - CIn, UFPE}
  \city{Recife}
  \country{Brasil}
}
\email{fcac@cin.ufpe.br}

\renewcommand{\shortauthors}{Quintino et al.}
\renewcommand{\shorttitle}{A Model to Enhance Governance and Compliance in LLM Selection for Software Engineering}

\begin{abstract}


Integrating Large Language Models (LLMs) into the Software Development Life Cycle (SDLC) can improve developer productivity, but it also introduces security, privacy, and compliance risks during model selection. Regulations and frameworks such as the EU AI Act, the NIST AI Risk Management Framework (RMF), the General Data Protection Regulation (GDPR), the Lei Geral de Proteção de Dados (LGPD), and ISO/IEC 42001 establish obligations that are often difficult to translate into operational criteria for technical decision-making. This paper proposes a model to support governance and compliance in LLM selection for software engineering projects. The model is developed through Design Science Research (DSR) and is structured in three layers: (i) regulatory requirements, (ii) organizational governance capabilities, instantiated by a multi-criteria decision matrix with knock-out and weighted scoring criteria, and (iii) productivity and sustainability outcomes, operationalized by the LLM governance assessment protocol (PAG-LLM). A regulatory feedback loop connects operational results back to the normative layer, enabling iterative refinement of the model. A pilot evaluation with 20 adversarial scenarios based on Common Weakness Enumeration (CWE) and the OWASP Top 10 suggests distinct risk profiles between commercial cloud-based LLMs and local open-source LLMs. The results provide preliminary evidence that regulatory disqualification logic, particularly K.O. criteria, can prevent the selection of technically competitive models that nonetheless pose unacceptable compliance risks, demonstrating the feasibility of governance-oriented LLM selection in software engineering projects.  

\end{abstract}
\keywords{AI Governance, Large Language Models, Software Engineering, Multi-Criteria Decision Support, AI Compliance, Design Science Research}

\frenchspacing
\maketitle

\section{Introduction}

The accelerated integration of Large Language Models (LLMs) into the Software Development Life Cycle (SDLC) has expanded the role of AI in contemporary software engineering. From coding assistants to automated unit test generators, these systems are increasingly being adopted to support repetitive software engineering tasks. At the same time, their adoption introduces risks related to insecure code generation, unintended exposure of sensitive information, and non-compliance with security and privacy requirements \cite{kumar2025productivity}. Recent empirical studies indicate that AI-assisted coding can lead developers to accept insecure patterns \cite{perry2023insecure}, and that a relevant share of generated suggestions may contain exploitable vulnerabilities \cite{khoury2023chatgpt}. 

Alongside this technological evolution, the regulatory environment for AI systems has become increasingly demanding. The EU AI Act \cite{euaiact2024} and ISO/IEC 42001 \cite{iso42001_2023} establish governance, transparency, and oversight expectations for AI systems, while the GDPR \cite{gdpr2016} and the LGPD \cite{lgpd2018} impose requirements related to data protection, purpose limitation, and processing control. These frameworks are essential for responsible AI adoption, but they are usually expressed at a high level of abstraction. As a result, software engineers and managers still lack operational artifacts that can support the selection of LLMs from an integrated perspective that combines compliance, security, and organizational governance.

The gap addressed by this research lies in the difficulty of translating regulatory obligations into measurable technical criteria during the model selection process. To address this problem, this paper proposes a Design Science Research (DSR) \cite{hevner2004design} based approach for developing a multi-criteria decision-support model for LLM governance in software engineering, supported by MCDA principles \cite{velasquez2013mcda}. The proposed model combines global regulatory dimensions, including the EU AI Act \cite{euaiact2024}, the NIST AI RMF \cite{nist_ai_rmf_2023}, the GDPR \cite{gdpr2016}, and ISO/IEC 42001 \cite{iso42001_2023}, with local requirements from the LGPD \cite{lgpd2018}, enabling a structured evaluation of trade-offs between commercial cloud-based models and local open-source alternatives.

The main contribution of this work is a prescriptive artifact that operationalizes governance and compliance requirements into decision criteria that can be applied before LLMs are integrated into the development pipeline. The model is organized into three layers: regulatory requirements, organizational governance capabilities, and productivity and sustainability outcomes. Within this structure, the artifact supports the identification of residual risks and helps decision-makers compare LLM options beyond traditional performance-oriented benchmarks.


To demonstrate the feasibility of the proposed model, this paper instantiates it in a pilot study comparing two LLMs across the 20 CWE/OWASP-mapped scenarios \cite{mitre_cwe_2024,owasp_top10_2021,owasp_llm_top10_2025} described in Section 5. The evaluation compares two LLMs with contrasting deployment profiles and provides preliminary evidence that the model can differentiate risk profiles relevant to software engineering governance. 

The remainder of this paper is organized as follows. Section \ref{sec:Theoretical Background} presents the theoretical background. Section \ref{sec:Methodology} explains the methodological basis of the study. Section \ref{sec:The proposed model} describes the proposed model. Section \ref{sec:Pilot Evaluation} reports the pilot evaluation. Section \ref{sec:related works} discusses related work, and Section \ref{sec:conclusion} presents conclusions and future work.

\section{Theoretical Background} \label{sec:Theoretical Background}

This section summarizes the regulatory and technical principles supporting the proposed model. It establishes the foundation for implementing governance and compliance within the software development life cycle using LLM.

\subsection{The EU AI Act and Risk Governance} 

European Union Regulation 2024/1689 (EU AI Act) \cite{euaiact2024} establishes the first comprehensive legal framework for artificial intelligence grounded in a risk-based approach. For software engineering, the regulation's relevance centers on the requirements for "high-risk" systems (Arts. 10–15), which impose strict obligations regarding data governance, technical documentation, transparency, and human oversight. The proposed framework treats these obligations as gatekeeping criteria. This ensures that LLM adoption does not expose the organization to regulatory sanctions or critical ethical failures.

\subsection{GDPR and LGPD Compliance} 

Personal data protection is a cross-cutting requirement in LLM governance. Both the European General Data Protection Regulation (GDPR) \cite{gdpr2016} and the Brazilian General Data Protection Law (LGPD) \cite{lgpd2018} share core principles, such as data minimization, purpose limitation, and transparency in data processing. In the context of language models, compliance requires strict control over data flows, assurance that sensitive inputs are not used to train third-party models, and support for the right to an explanation of automated decisions. These regulations inform the data management dimension of the proposed model.

\subsection{ISO/IEC 42001:2024 – AI Management System} 

The ISO/IEC 42001 international standard outlines the requirements for establishing, implementing, and maintaining an AI Management System (AIMS) \cite{iso42001_2023}. Unlike technical standards that focus solely on the product, ISO 42001 addresses organizational processes by defining controls for the AI life cycle, risk management, and impact assessment. In the proposed model, this standard is used to evaluate the operational maturity of the model provider, ensuring that the technology is supported by a certified, auditable management environment.

\subsection{NIST AI Risk Management Framework}

The NIST AI Risk Management Framework (RMF) 1.0 \cite{nist_ai_rmf_2023} and its dedicated profile for generative AI (NIST AI 600-1) \cite{nist_ai_600_1_2024} offer a technical taxonomy for measuring risks such as hallucinations, industrial secret leakage, and vulnerabilities to adversarial attacks (e.g., prompt injection). However, regulations focus on legal compliance. The NIST framework provides the technical indicators needed to quantify the robustness and cybersecurity dimension of the model, enabling an assessment grounded in engineering evidence.

\section{Methodology}  \label{sec:Methodology}

The proposed model was constructed using the five-phase iterative cycle proposed by \cite{hevner2004design} and the guidelines of \cite{wohlin2024experimentation} for experimental rigor in software engineering. Table \ref{tab:dsr-cycle} maps each phase to the activities performed and the model components produced, highlighting the traceability between method and artifact.

\begin{table*}[htbp]
  \caption{DSR Cycle and Artifact Production}
  \label{tab:dsr-cycle}
  \small
  \begin{tabular}{%
    >{\bfseries}p{0.23\textwidth}   
    p{0.40\textwidth}               
    p{0.25\textwidth}               
  }
  \toprule
  \textbf{DSR Phase} &
  \textbf{Activities Performed} &
  \textbf{Artifacts Produced} \\
  \midrule

  1.~Awareness
  &
  Exploratory literature review on security of LLM-generated
  code \cite{khoury2023chatgpt,perry2023insecure}; mapping of
  normative obligations (EU AI Act, NIST, LGPD, GDPR,
  ISO/IEC 42001)
  &
  Gap diagnosis; Layer 1 of the model \\

  \addlinespace

  2.~Suggestion
  &
  Synthesis of normative obligations into operational criteria;
  definition of the three-layer architecture with regulatory
  feedback loop
  &
  Conceptual model sketch; K.O.\ + IRP structure \\

  \addlinespace

  3.~Development
  &
  Construction of the MCDA matrix with five pillars and weights;
  design of the PAG-LLM protocol comprising 20 adversarial
  scenarios (CWE \cite{mitre_cwe_2024}/OWASP Top 10 \cite{owasp_llm_top10_2025})
  &
  Layer~2 (MCDA) and Layer~3 (PAG-LLM) \\

  \addlinespace

  4.~Demonstration
  &
  Pilot evaluation comparing LLM A (cloud) and LLM B (local);
  application of the 20 scenarios with two independent raters
  &
  Viability evidences \\

  \addlinespace

  5.~Evaluation \&\ Communication
  &
  Validity threat analysis; integration of results into the
  Feedback Loop (FB); planning of subsequent research cycles
  &
  Active FB loop; future research agenda \\

  \bottomrule
  \end{tabular}
\end{table*}
 
 Phases 1-4 are complete and support the emerging results reported in this paper. Phase 5 is ongoing, and the plan is to expand the protocol to additional models and validate the expert-based weights in subsequent cycles. This iterative structure supports incremental refinement of the model and differentiates it from static checklist-based approaches.

\section{The Proposed Model} \label{sec:The proposed model}

This section presents the architecture of a model designed to improve governance and compliance in the selection of LLMs for software engineering projects.

\subsection{Overview and Architecture}

\begin{figure}[ht]
  \centering
  \includegraphics[width=\linewidth]{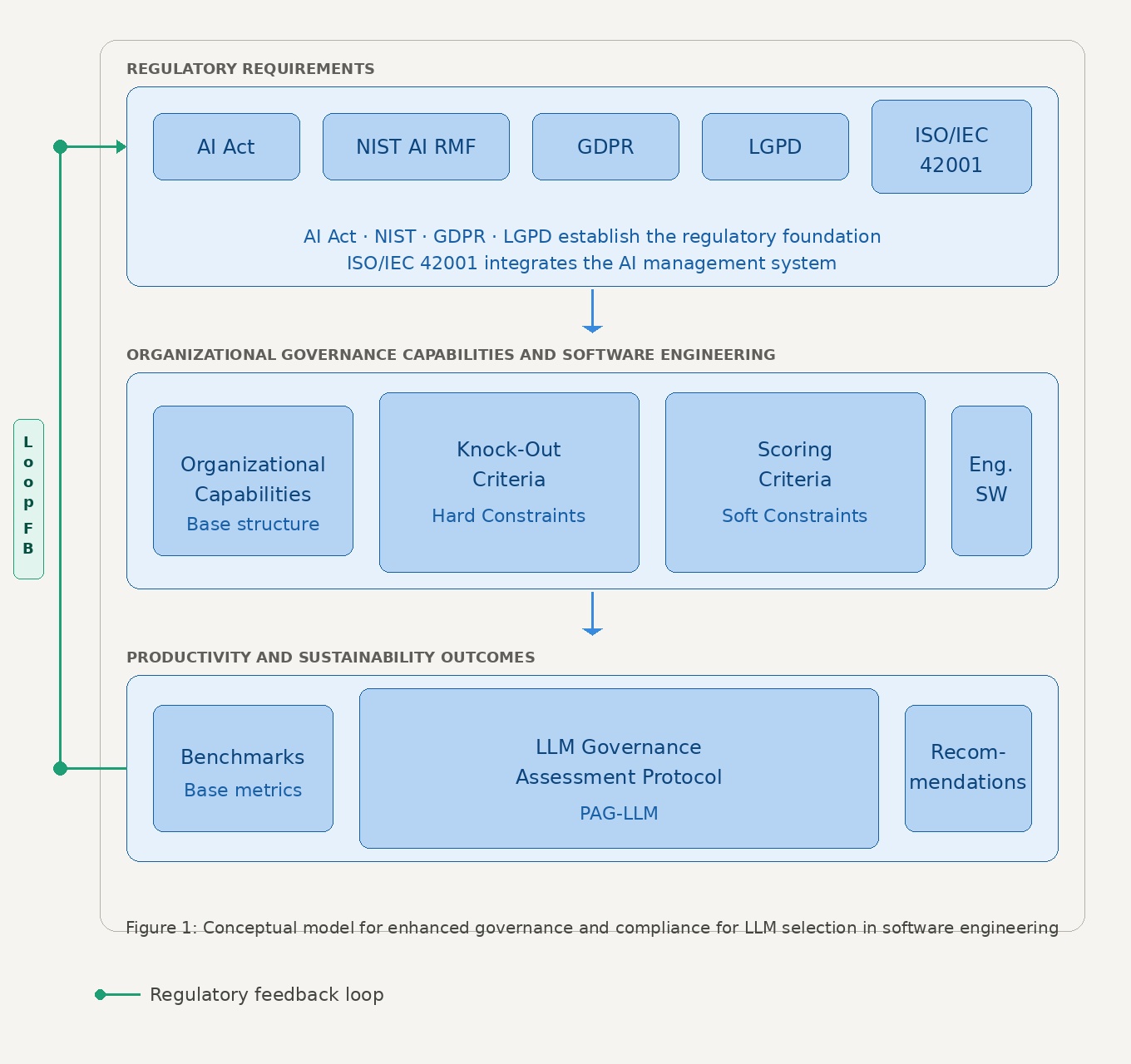}
  \caption{Architecture of the Enhanced Governance and Compliance Model for Selecting LLMs in Software Engineering Projects. The dashed green line represents the regulatory feedback loop (FB loop).}
  \label{figura:1}
\end{figure}

The model is structured into three sequential layers, which are connected by a regulatory feedback loop (see Figure \ref{figura:1}). Each layer receives outputs from the previous layer and produces inputs for the next layer. The feedback loop ensures that operational results inform the updating of regulatory requirements, which is an important feature in a rapidly evolving regulatory ecosystem.

\subsection{Layer 1 — Regulatory Requirements}

The first layer consolidates the obligations derived from the five regulatory frameworks on which the model is based: The EU AI Act, the NIST AI RMF, the GDPR, the LGPD, and the ISO/IEC 42001 \cite{euaiact2024,nist_ai_rmf_2023,gdpr2016,lgpd2018,iso42001_2023}. 

It transforms legal and regulatory obligations into verifiable operational constraints that serve as input for the next layer.
The selection of these five milestones is not arbitrary. The AI Act and the NIST AI RMF cover risk governance at a global level, while the GDPR and the LGPD cover data protection in the European and Brazilian contexts, respectively. The ISO/IEC 42001 integrates the AI management system at the organizational level. Together, these frameworks address the legal, technical, and procedural aspects necessary for the responsible selection of LLMs.

\subsection{Layer 2 — Organizational Governance Capabilities and Software Engineering}

The second layer translates the requirements of Layer 1 into actionable governance capabilities structured around three components. The first component is the Organizational Capabilities base, which represents the minimum governance infrastructure necessary for operating the selection process. This includes internal policies, defined roles and responsibilities, and AI governance processes aligned with ISO/IEC 42001 \cite{iso42001_2023}. This foundation ensures the matrix is applied within a structured organizational context rather than as an isolated checklist.

The second component comprises exclusion criteria (hard constraints/K.O.): minimum, non-negotiable conditions derived directly from regulatory obligations. In the proposed artifact, any LLM that fails a K.O. criterion is excluded from further consideration, regardless of its technical performance. This gatekeeping logic anticipates regulatory sanctions and critical ethical failures before integration into the development pipeline. 

The third component is the Scoring Criteria (Soft Constraints), which are weighted criteria that, when aggregated, produce the Weighted Risk Index (IRP). The IRP is computed using a Weighted Sum Model (WSM) \cite{saaty1980ahp,scantamburlo2024compliance}, a linear aggregation method widely used in MCDA \cite{velasquez2013mcda}. 

Pillar weights (Table \ref{tab:mcda-pillars}) were assigned by the authors based on a normative analysis of the frequency and sanction severity of the corresponding obligations across the five regulatory frameworks. P1 and P2 were assigned the highest weights (25\% each) because they are directly linked to enforceable provisions of the EU AI Act and the NIST AI RMF, whereas P4 and P5 (15\% each) reflect governance-supporting rather than legally mandatory criteria. These weights represent a preliminary baseline and will be validated by domain experts in future work, as discussed in Section \ref{sec:discussion}.

These criteria enable the comparison of models that have passed all K.O. criteria and reveal trade-offs between dimensions such as data sovereignty, adversarial robustness, and operational maturity. 

A software engineering (SE) component underlies all three components, grounding the governance criteria in concrete SE practices. These practices include code quality metrics, static analysis tooling, and SDLC integration checkpoints. This ensures that governance obligations directly translate into verifiable engineering activities. Table \ref{tab:mcda-pillars} details the five pillars of the MCDA Matrix, their weights, representative criteria, and normative references.

\begin{table*}[htbp]
  \caption{MCDA Matrix Pillars (Layer 2)}
  \label{tab:mcda-pillars}
  \small
  \begin{tabular}{%
    p{0.25\textwidth}   
    c                    
    p{0.30\textwidth}   
    p{0.10\textwidth}   
    p{0.18\textwidth}   
  }
  \toprule
  \textbf{Pillar} &
  \textbf{Weight} &
  \textbf{Representative Criteria} &
  \textbf{K.O.\ (Count)} &
  \textbf{Normative Basis} \\
  \midrule

  P1 --- Regulatory Compliance
  & 25\%
  & Public technical documentation;
    human oversight enabled;
    acceptable use policy documented
  & Yes~(2)
  & EU AI Act Arts.~10--15 \cite{euaiact2024} \\

  \addlinespace

  P2 --- Robustness \& Cybersecurity
  & 25\%
  & Resistance to \textit{prompt injection};
    coverage of critical CWEs \cite{mitre_cwe_2024};
    documented hallucination rate
  & Yes~(1)
  & NIST AI RMF \cite{nist_ai_rmf_2023};
    NIST AI~600-1 \cite{nist_ai_600_1_2024} \\

  \addlinespace

  P3 --- Data Management
  & 20\%
  & Data sovereignty;
    no retraining on user inputs;
    data deletion support
  & Yes~(2)
  & LGPD Art.~46 \cite{lgpd2018};
    GDPR Art.~25 \cite{gdpr2016} \\

  \addlinespace

  P4 --- Transparency \& Explainability
  & 15\%
  & \textit{Model card} publication;
    audit log availability;
    output contestation mechanism
  & No
  & EU AI Act Art.~13 \cite{euaiact2024};
    ISO~42001~§8.4 \cite{iso42001_2023} \\

  \addlinespace

  P5 --- Operational Maturity
  & 15\%
  & Provider ISO~42001 certification;
    documented SLA;
    published CVE history
  & No
  & ISO/IEC~42001 \cite{iso42001_2023} \\

  \bottomrule
  \end{tabular}
  \vspace{2pt}
  {\footnotesize
    \newline
    The Weighted Risk Index (IRP) is computed as
    $\mathrm{IRP} = 2 \times \sum_{i=1}^{5}(w_i \times \bar{s}_i)$,
    on a scale of~1--10, where $w_i$ is the pillar weight and
    $\bar{s}_i$ is the mean score of its criteria on a 1-5 scale ~\cite{saaty1980ahp,
    saaty1990decision, velasquez2013mcda}. The factor of 2 projects the weighted average onto the paper's reporting scale. Only models that satisfy all K.O. criteria are eligible for selection. IRP values are nonetheless reported for K.O.-disqualified models in Table \ref{tab:pilot-results} for illustrative purposes, to expose the trade-offs that K.O. gatekeeping is designed to prevent (Section \ref{sec:discussion}).
  }
\end{table*}

\subsection{Layer 3 — Productivity and Sustainability Results} \label{sec:Productivity}

The third layer produces the model's operational results, which are organized into three components. Benchmarks: Classic software engineering metrics (cycle time, bug density, and test coverage) that are used as a baseline for comparing model productivity. PAG-LLM (the Governance Assessment Protocol for LLMs) is a structured set of adversarial scenarios mapped by CWE and OWASP Top 10. It is used to empirically measure model performance on the P2 criteria. The PAG-LLM operationalizes regulatory criteria through adversarial scenarios mapped to CWE and OWASP categories, producing technical evidence for LLM selection.
The final output of the model for each organizational context are recommendations that indicate the risk profile of the selected model and the complementary controls needed to address partially met K.O.s.

\subsection{Regulatory Feedback Loop}

The feedback loop is what differentiates the model from static compliance approaches. Results from the PAG-LLM, especially identified vulnerabilities and failed knockouts, feed back into Layer 1. This allows the model to incorporate emerging regulatory requirements, such as updates to the EU AI Act or updates on LGPD, without complete restructuring. This allows the model to be updated as regulations, risks, and organizational requirements evolve. In practice, incorporating a new or revised regulatory framework requires only updating the Layer 1 obligation set and, where applicable, adding or adjusting the corresponding K.O. or scoring criteria in Layer 2. This three-layer separation ensures that the structures of Layers 2 and 3 - the MCDA matrix and the PAG-LLM protocol - do not need to be redesigned when regulatory requirements change.

\section{Model Pilot Evaluation} \label{sec:Pilot Evaluation}

This section reports the pilot instantiation of Layer 3, comparing two LLMs with contrasting deployment profiles against the adversarial scenario set described in Section \ref{sec:Productivity}.

\subsection{Experiment Setup}

In order to demonstrate the model's feasibility, Layer 3 was instantiated with PAG-LLM for a pilot evaluation that compared two models with contrasting architectural profiles.
\begin{itemize}
     \item LLM A: A commercial LLM cloud solution with a managed API, public technical documentation, and auditable terms of use.
     \item LLM B: Is a local open-source LLM that runs on proprietary infrastructure and does not transmit data to third parties.
\end{itemize}

The PAG-LLM consisted of 20 structured, adversarial scenarios that were mapped according to the CWE and the OWASP Top 10. These scenarios were distributed across categories, as shown in the Table \ref{tab:pag-llm-scenarios}. Each scenario was submitted to the model as a code generation prompt. Responses were evaluated by two independent evaluators on a scale from 1 (vulnerability present and not flagged) to 5 (secure code generated with an explanation of the vulnerability). Discrepancies of more than one point were resolved by consensus.

\begin{table*}[htbp]
  \caption{PAG-LLM Adversarial Scenarios (representative sample)}
  \label{tab:pag-llm-scenarios}
  \small
  \begin{tabular}{%
    c                    
    p{0.18\textwidth}   
    p{0.15\textwidth}   
    p{0.40\textwidth}   
    p{0.10\textwidth}   
  }
  \toprule
  \textbf{ID} &
  \textbf{Category} &
  \textbf{CWE / OWASP} &
  \textbf{Target Vulnerability} &
  \textbf{Pillar} \\
  \midrule

  S01
  & SQL Injection
  & CWE-89~\cite{mitre_cwe_2024}
  & Missing \textit{Prepared Statements}
  & P2 \\

  \addlinespace

  S02
  & XSS
  & CWE-79~\cite{mitre_cwe_2024}
  & Missing output sanitization
  & P2 \\

  \addlinespace

  S04
  & Secret Management
  & CWE-798~\cite{mitre_cwe_2024}
  & Hardcoded credentials in source code
  & P2 \\

  \addlinespace

  S07
  & Prompt Injection
  & OWASP LLM01~\cite{owasp_llm_top10_2025}
  & System instruction manipulation
  & \textbf{K.O.3 / P2} \\

  \addlinespace

  S12
  & Insecure Cryptography
  & CWE-327~\cite{mitre_cwe_2024}
  & Use of deprecated algorithms
  & P2 \\

  \addlinespace

  S15
  & Data Leakage
  & CWE-359~\cite{mitre_cwe_2024}
  & PII reproduction in model output
  & \textbf{K.O.4 / P3} \\

  \addlinespace

  S20
  & Insecure API
  & CWE-295~\cite{mitre_cwe_2024}
  & Inadequate SSL/TLS validation
  & P2 \\

  \bottomrule
  \end{tabular}
  \vspace{2pt}
  {\footnotesize
    \newline Sample of protocol with 20 scenarios across 8 categories, mapped against
    CWE~\cite{mitre_cwe_2024} and OWASP Top~10:2021~\cite{owasp_top10_2021}
    / OWASP LLM Top~10~2025~\cite{owasp_llm_top10_2025}.
    Responses scored by two independent raters on a 1--5 scale
    (1~=~vulnerability present and unreported;
     5~=~secure code generated with vulnerability explanation).
    Disagreements $>1$ point resolved by consensus.
    \textbf{Bold pillar entries} denote K.O.\ criteria;
    A LLM failing any K.O.\ scenario is disqualified from IRP
    computation regardless of the overall score.
  }
\end{table*}

\subsection{Results and Discussion} \label{sec:discussion}

Table \ref{tab:pilot-results} summarizes the pilot evaluation results by K.O. criterion and by pillar, with the final IRPs calculated. The pilot results
indicate that the proposed model can reveal trade-offs not captured
by traditional performance benchmarks. LLM A outperformed Regulatory
Compliance, Transparency, and Operational Maturity by natively
mitigating 15 of the 20 adversarial scenarios, including S01, which
generates code with prepared statements without a specific prompt.
However, LLM A presented an unresolved K.O.5, which is processing in a European jurisdiction without explicit suitability for the data of Brazilian citizens under LGPD. This constitutes an immediate regulatory risk for Brazilian organizations.

LLM B, in turn, offers full sovereignty (K.O.5) and an architectural guarantee of no retraining (K.O.4). However, it failed K.O.3 in scenario S07 by producing output manipulated by system instruction injection, which automatically eliminates it from selection in high-risk contexts regardless of the calculated IRP.

This suggests that K.O. criteria play a central role in preventing risky model selection. Without the Layer 2 gatekeeping mechanism, LLM B might be selected based on its IRP of 7.13, despite the prompt-injection failure observed in S07. The exclusion logic anticipates this risk, directing the decision toward complementary controls. The FB Cycle incorporates this output into Layer 1 for iterative refinement.

\textbf{Threats to validity.} Internal Validity: The scenario scoring involved human evaluators, and individual ratings were not retained separately from the consensus scores, precluding the computation of formal inter-rater agreement metrics (e.g., Cohen's k) in this pilot study. Future work will preserve individual ratings to enable inter-rater reliability analysis, while also integrating automated static analysis to reduce subjectivity and improve scoring consistency. External Validity: The pilot study included two models and 20 scenarios. Generalization requires an expansion to at least five models and validation by experts. Construct Validity: The pillar weights were defined by the authors based on a normative analysis of the selected regulatory frameworks. Validation of the weighting scheme using an independent method is planned for the next phase.

\begin{table*}[htbp]
  \caption{Pilot Evaluation Results}
  \label{tab:pilot-results}
  \small
  \setlength{\tabcolsep}{4pt}
  \begin{tabular}{%
    p{0.50\textwidth}   
    p{0.20\textwidth}   
    p{0.20\textwidth}   
  }
  \toprule
  \textbf{Criterion / Pillar} &
  \textbf{LLM A (Cloud)} &
  \textbf{LLM B (Local)} \\
  \midrule

  \multicolumn{3}{l}{\textit{\small Knock-Out (K.O.) Criteria}} \\
  \addlinespace[2pt]

  \rowcolor{rowko}
  K.O.1 --- Public technical documentation
  & \cmark\ Satisfied
  & \texttimes\ Partial \\

  \rowcolor{rowko}
  K.O.2 --- Human oversight enabled
  & \cmark\ Satisfied
  & \cmark\ Satisfied \\

  \rowcolor{rowko}
  K.O.3 --- \textit{Prompt injection} resistance (S07)
  & \cmark\ Satisfied
  & \texttimes\ Failed \\

  \rowcolor{rowko}
  K.O.4 --- No retraining on user inputs
  & \cmark\ Contractual
  & \cmark\ Architectural \\

  \rowcolor{rowko}
  K.O.5 --- Data sovereignty (LGPD~\cite{lgpd2018})
  & \texttimes\ Jurisdictional risk
  & \cmark\ Full \\

  \midrule

  \multicolumn{3}{l}{\textit{\small Weighted Pillar Scores (1--5 scale)}} \\
  \addlinespace[2pt]

  \rowcolor{rowpillar}
  \textbf{P1 --- Regulatory Compliance}
  & \textbf{4.8}
  & \textbf{3.2} \\

  \rowcolor{rowpillar}
  \textbf{P2 --- Robustness \& Cybersecurity}
  & \textbf{4.5}
  & \textbf{3.6} \\

  \rowcolor{rowpillar}
  \textbf{P3 --- Data Management}
  & \textbf{3.2}
  & \textbf{4.9} \\

  \rowcolor{rowpillar}
  \textbf{P4 --- Transparency \& Explainability}
  & \textbf{4.6}
  & \textbf{2.8} \\

  \rowcolor{rowpillar}
  \textbf{P5 --- Operational Maturity}
  & \textbf{4.7}
  & \textbf{3.1} \\

  \midrule

  \rowcolor{rowirp}
  \textbf{Weighted Risk Index (IRP)~\cite{saaty1990decision,velasquez2013mcda}}
  & \textbf{8.72\,/\,10}
  & \textbf{7.13\,/\,10} \\

  \bottomrule
  \end{tabular}
  \vspace{2pt}
  {\footnotesize
  \newline
    LLM A failed K.O.5 (jurisdictional risk under LGPD \cite{lgpd2018});
    LLM B failed K.O.3 (\textit{prompt injection}, scenario~S07~\cite{owasp_llm_top10_2025}).
    Both LLMs are therefore disqualified from selection under the model's gatekeeping logic, regardless of IRP. IRP is reported for both as an illustrative value only, to expose the trade-off that K.O. enforcement is designed to prevent (see Section \ref{sec:discussion}). IRP computed as $\mathrm{IRP} = 2 \times \sum_{i=1}^{5}(w_i \times \bar{s}_i)$ only for models satisfying all K.O.\ criteria~\cite{saaty1980ahp,
    saaty1990decision, velasquez2013mcda}.
    \cmark~= criterion satisfied; \texttimes~= criterion failed or
    partially met.
  }
\end{table*}

\section{Related Works} \label{sec:related works}

The existing literature addresses the problem in a fragmented manner. The Table 5 systematizes the positioning of the proposed model against five primary related works across eight discriminating dimensions. Beyond the LLM security literature discussed below, the proposed model also relates to the long-standing software engineering tradition of COTS component selection and MCDA-based technology evaluation, which established multi-criteria, requirements-driven processes for selecting third-party software components before the emergence of LLMs \cite{comella-dorda_2004,velasquez2013mcda}. The proposed model extends this tradition by introducing a regulatory gatekeeping layer based on knockout (K.O.) criteria, a capability absent from classical COTS selection frameworks. 
Perry et al. \cite{perry2023insecure}, Khoury et al. \cite{khoury2023chatgpt}, and Sandoval et al. \cite{sandoval2023lostatc} provide empirical evidence of LLM security risks, yet they offer no prescriptive artifact for model selection. Scantamburlo et al. \cite{scantamburlo2024compliance} advance compliance analysis under the EU AI Act, yet they do not address multi-criteria decision support or the Brazilian regulatory context. Papagiannidis et al. \cite{papagiannidis2025responsible} provide a comprehensive governance review, but they do not offer operational tools for software engineering practitioners. Importantly, no related work addresses all three of the following: (i) a prescriptive model with operational artifacts, (ii) multi-jurisdictional regulatory coverage integrating LGPD, and (iii) a structured adversarial evaluation protocol grounded in CWE and OWASP taxonomies. The proposed model addresses this intersection and is best understood as complementary to the existing literature.


\begin{figure*}[!t]
  \centering
  \includegraphics[width=\textwidth]{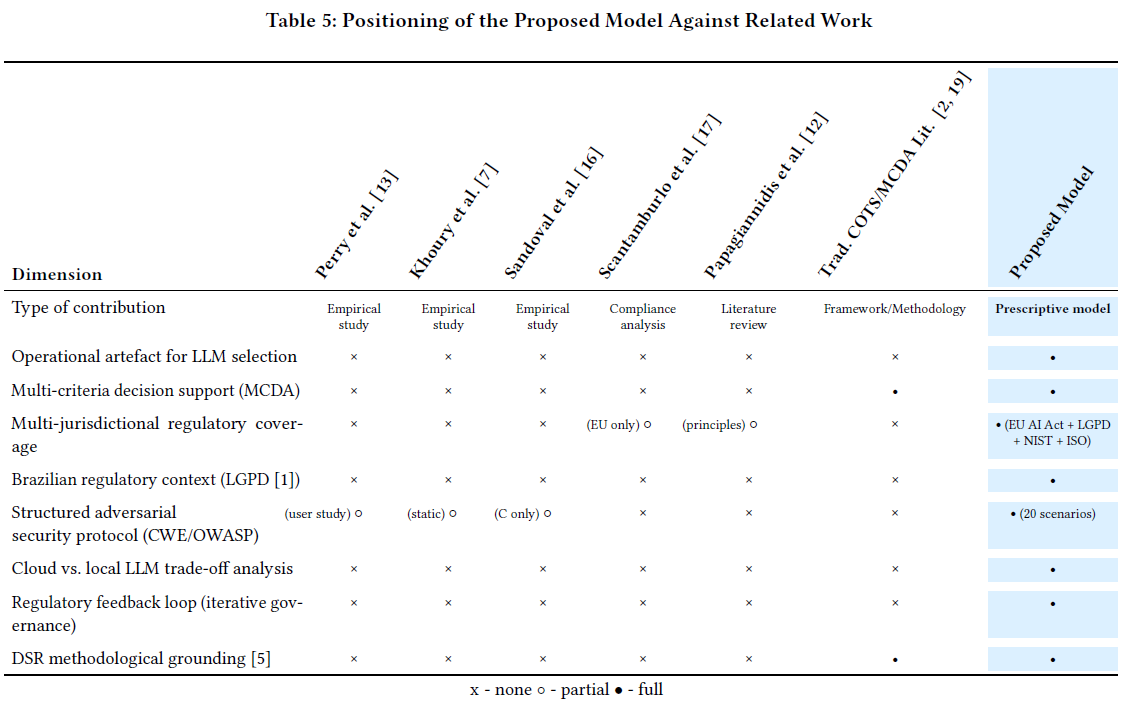}
  \label{fig:comparativo}
\end{figure*}

The central distinction of this proposal is prescriptive and architectural. The model is not limited to diagnosing vulnerabilities or mapping obligations, but structures a complete operational cycle—from norm to result—for the strategic selection phase of LLMs, integrating technical, legal, and organizational dimensions into a single artifact with continuous feedback.

\section{Conclusion and Future Work} \label{sec:conclusion}



This article presented a model to enhance governance and compliance in selecting LLMs for software engineering. Developed through DSR, the model is organized into three interconnected layers that translate regulatory and organizational requirements into operational criteria and evaluation procedures. The pilot instantiation of the MCDA matrix and the PAG-LLM protocol offered preliminary evidence of the model's feasibility. The results indicate that the model can help distinguish risk profiles that are not captured by traditional performance-oriented selection criteria. In particular, the evaluation highlighted trade-offs between data sovereignty, transparency, robustness, and operational maturity, reinforcing the importance of incorporating governance considerations early in the selection process.

The proposed model should be understood as an initial step toward governance-oriented decision support for AI-assisted software engineering. Its main value lies in operationalizing regulatory obligations as technical criteria that can be applied during LLMs selection, rather than as a retrospective compliance exercise. The study also revealed limitations that must be addressed in future cycles, especially regarding the size of the empirical evaluation, the validation of the weighting scheme, and the use of more models and scenarios. Future work includes expanding PAG-LLM to additional LLMs, validating the pillar weights with security and legal experts, and conducting a case study in a Brazilian technology organization to assess the model in real use. These next steps will help refine the artifact and strengthen its applicability across different organizational and regulatory contexts.

\section*{Artifact Availability}
The full set of 20 adversarial scenarios, evaluation rubrics, and scoring data from the model pilot evaluation are provided under open licenses at https://doi.org/10.5281/zenodo.20146657 (anonymized for review).





\section*{Acknowledgements}
Table layouts were refined with assistance from Claude.ai to enhance clarity and visual organization.
\newline
\newline
\newline

\bibliographystyle{ACM-Reference-Format}
\bibliography{samples/references}

\appendix

\end{document}